\documentclass{aa}
\usepackage{graphicx}

\def\jh{\mbox{$(J-H)$}}
\def\hk{\mbox{$(H-K_s)$}}
\def\mMJ{\mbox{$(m-M)_J$}}
\def\mMo{\mbox{$(m-M)_O$}}
\def\ebv{\mbox{$E(B-V)$}}
\def\ejh{\mbox{$E(J-H)$}}
\def\ehk{\mbox{$E(H-K_s)$}}
\def\rc{\mbox{$R_{\rm core}$}}
\def\rl{\mbox{$R_{\rm lim}$}}
\def\ms{\mbox{$M_\odot$}}
\def\ds{\mbox{$d_\odot$}}
\def\dgc{\mbox{$d_{\rm GC}$}}
\def\jj{\mbox{$J$}}
\def\hh{\mbox{$H$}}
\def\ks{\mbox{$K_s$}}

\def\mobs{\mbox{$m_{\rm obs}$}}

\def\tr{\mbox{$t_{\rm rel}$}}

\begin{document}

\title{Faint open clusters with 2MASS: BH\,63, Lyng\aa\,2, Lyng\aa\,12 and King\,20}

\author{E. Bica\inst{1} \and C. Bonatto\inst{1} \and R. Blumberg \inst{1}}

\offprints{Ch. Bonatto}

\institute{Universidade Federal do Rio Grande do Sul, Instituto de F\'\i sica, 
CP\,15051, Porto Alegre 91501-970, RS, Brazil\\
\email{charles@if.ufrgs.br, bica@if.ufrgs.br, blumberg@if.ufrgs.br}
\mail{charles@if.ufrgs.br} }

\date{Received --; accepted --}

\abstract{Structural and dynamical parameters of faint open clusters are probed with 
quality 2MASS-photometry and analytical procedures developed for bright clusters.}
{We derive fundamental parameters of the faint open clusters Lyng\aa\,2, BH\,63, 
Lyng\aa\,12 and King\,20, the last three of which have no prior determinations. We 
also focus on the structure and dynamical state of these clusters.}
{\jj, \hh\ and \ks\ 2MASS photometry with errors smaller than 0.2\,mag are used to build CMDs, 
radial density profiles, colour-colour diagrams, luminosity and mass functions. Colour-magnitude 
filters are used to isolate probable member stars. Field-star decontamination is applied to 
Lyng\aa\,2, Lyng\aa\,12 and King\,20.}
{Reddening values are in the range $0.22\leq\ebv\leq1.9$, with BH\,63 the most reddened object. 
Ages of Lyng\aa\,2, King\,20, Lyng\aa\,12 and BH\,63 are $\approx90$, $\approx200$, $\approx560$ and 
$\approx700$\,Myr, respectively. The radial density distributions of Lyng\aa\,12 and King\,20 are 
well-represented 
by King profiles. Lyng\aa\,2 and BH\,63 are very small with core and limiting radii of $\approx0.12$\,pc 
and $\approx1.5$\,pc. Yet, they fit in the small-radii tail of the open cluster size distribution. 
Lyng\aa\,12 and King\,20 have $\rc\approx0.43$\,pc and $\rl\approx3.9$\,pc. Lyng\aa\,2 and Lyng\aa\,12 
are inside the Solar circle. Total stellar masses (extrapolating the MFs to stars with 0.08\,\ms) range 
from $\approx340$\,\ms\ (BH\,63) to $\approx2300$\,\ms\ (Lyng\aa\,12). Observed masses are $\sim1/4$ 
of these values. In all clusters the core mass function is flatter than the halo's.}
{Faint open clusters can be probed with 2MASS when associated with colour-magnitude filters and
field-star decontamination. BH\,63 appears to be in an advanced dynamical state, both in the core 
and halo. To a lesser degree the same applies to King\,20. Marginal evidence of dynamical evolution is 
present in the cores of Lyng\aa\,2 and Lyng\aa\,12.}


\titlerunning{Parameters of faint open clusters}

\authorrunning{E. Bica et al.}

\maketitle

\section{Introduction}
\label{intro}

Because it is relatively simple to estimate ages and distances of open clusters (OCs), they have
become fundamental probes of Galactic disc properties (Lyng\aa~\cite{Ly82}; Janes \& Phelps 
\cite{JP94}; Friel \cite{Friel95}; Bonatto et al. \cite{BKBS06}; Piskunov et al. \cite{Pi2006}). 
However, the proximity of most of the OCs to the plane and the corresponding high values of 
reddening and field-star contamination (Bonatto et al. \cite{BKBS06} and references therein) 
usually restrict this analysis to the more populous OCs and/or those located a few kpc from 
the Sun. 

To probe disc structure Bonatto et al. (\cite{BKBS06}) employed 654 OCs with fundamental parameters 
(such as age, distance from the Sun, and reddening) mostly from the WEBDA\footnote{\em 
http://obswww.unige.ch/webda} database (Mermilliod \cite{Merm1996}). They found that a large fraction 
of the faint and/or poorly populated OCs must be overwhelmed in the field, particularly in bulge/disc 
directions. Piskunov et al. (\cite{Pi2006}) used 650 OCs from ASCC-2.5 data to infer kinematical
similarities among OC groups. Both works estimate a total population of $\sim10^5$ OCs in the Galaxy. 
Dias et al. (\cite{Dias2002}) reported 1756 optically visible OCs and candidates in their catalogue. 
These works suggest that besides those not yet observed, there is a large number of OCs and candidates 
whose properties have not been explored.

\begin{figure*}
\begin{minipage}[b]{0.50\linewidth}
\includegraphics[width=\textwidth]{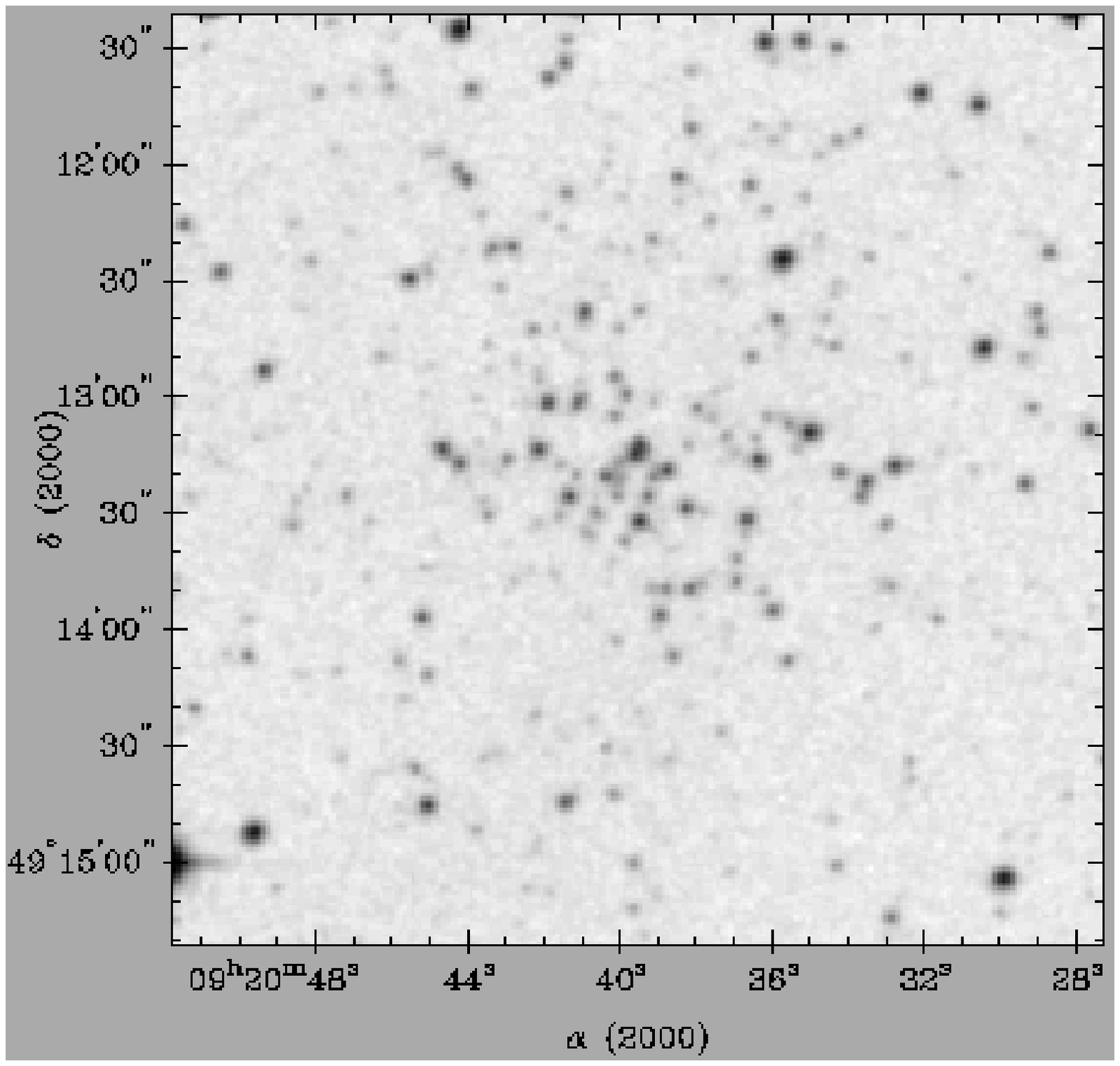}
\end{minipage}\hfill
\begin{minipage}[b]{0.50\linewidth}
\includegraphics[width=\textwidth]{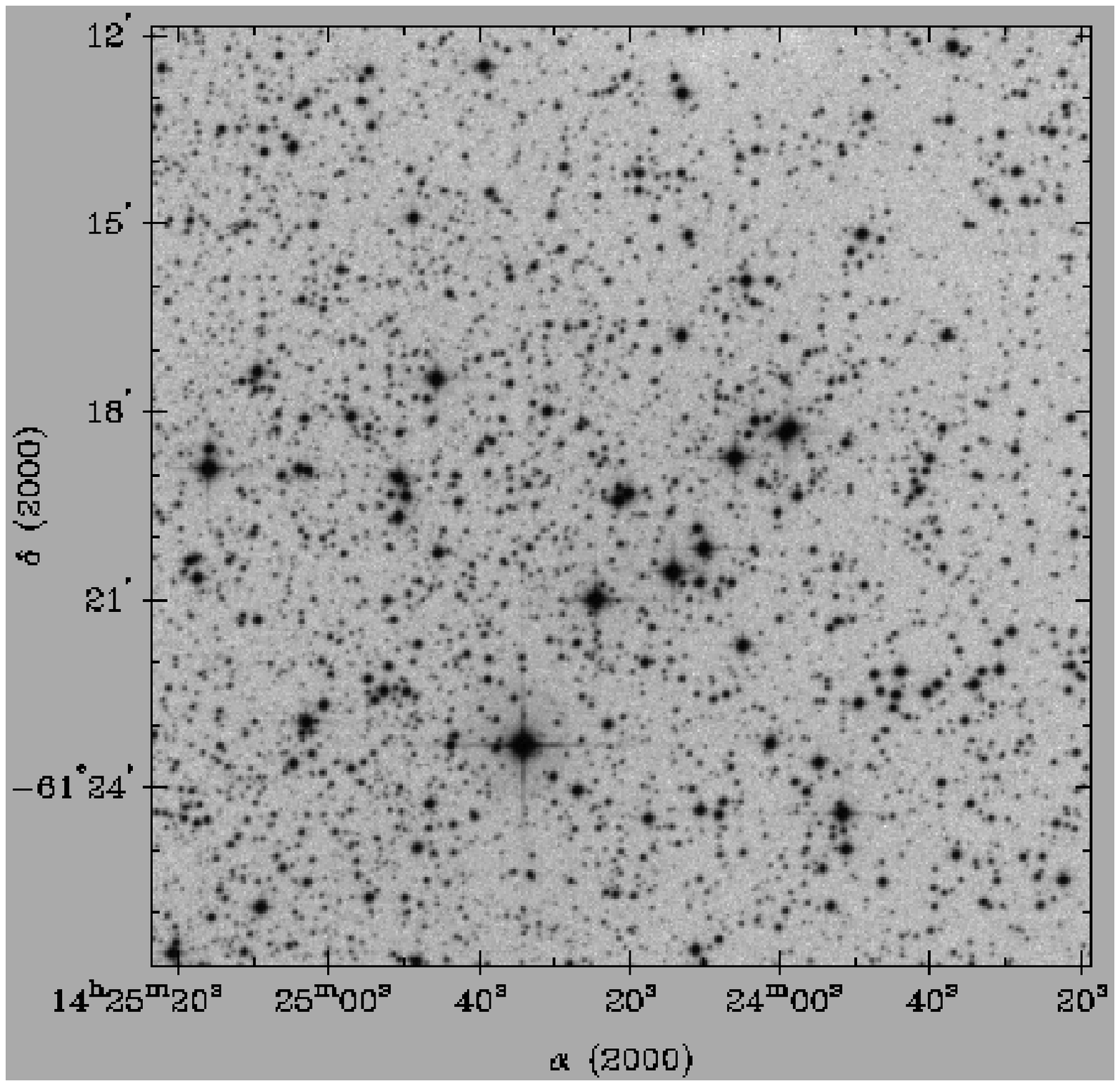}
\end{minipage}\hfill
\caption[]{Left panel: $4\arcmin\times4\arcmin$ XDSS R image of BH\,63. Right panel: $15\arcmin\times15\arcmin$
XDSS R image of Lyng\aa\,2. Images centered on the optimised coordinates (cols.~5 and 6 of Table~\ref{tab1}).}
\label{fig1}
\end{figure*}

\begin{figure*}
\begin{minipage}[b]{0.50\linewidth}
\includegraphics[width=\textwidth]{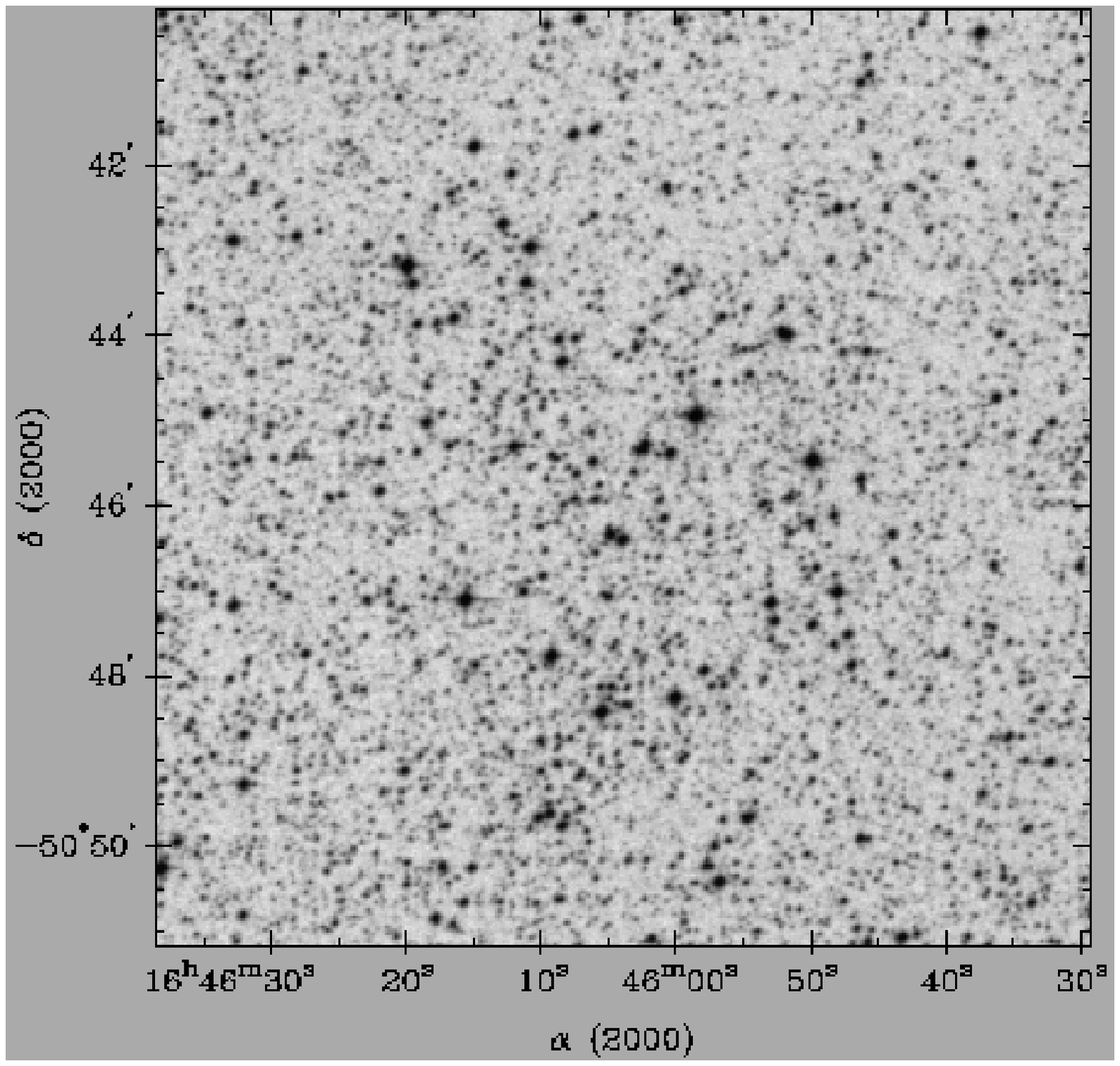}
\end{minipage}\hfill
\begin{minipage}[b]{0.50\linewidth}
\includegraphics[width=\textwidth]{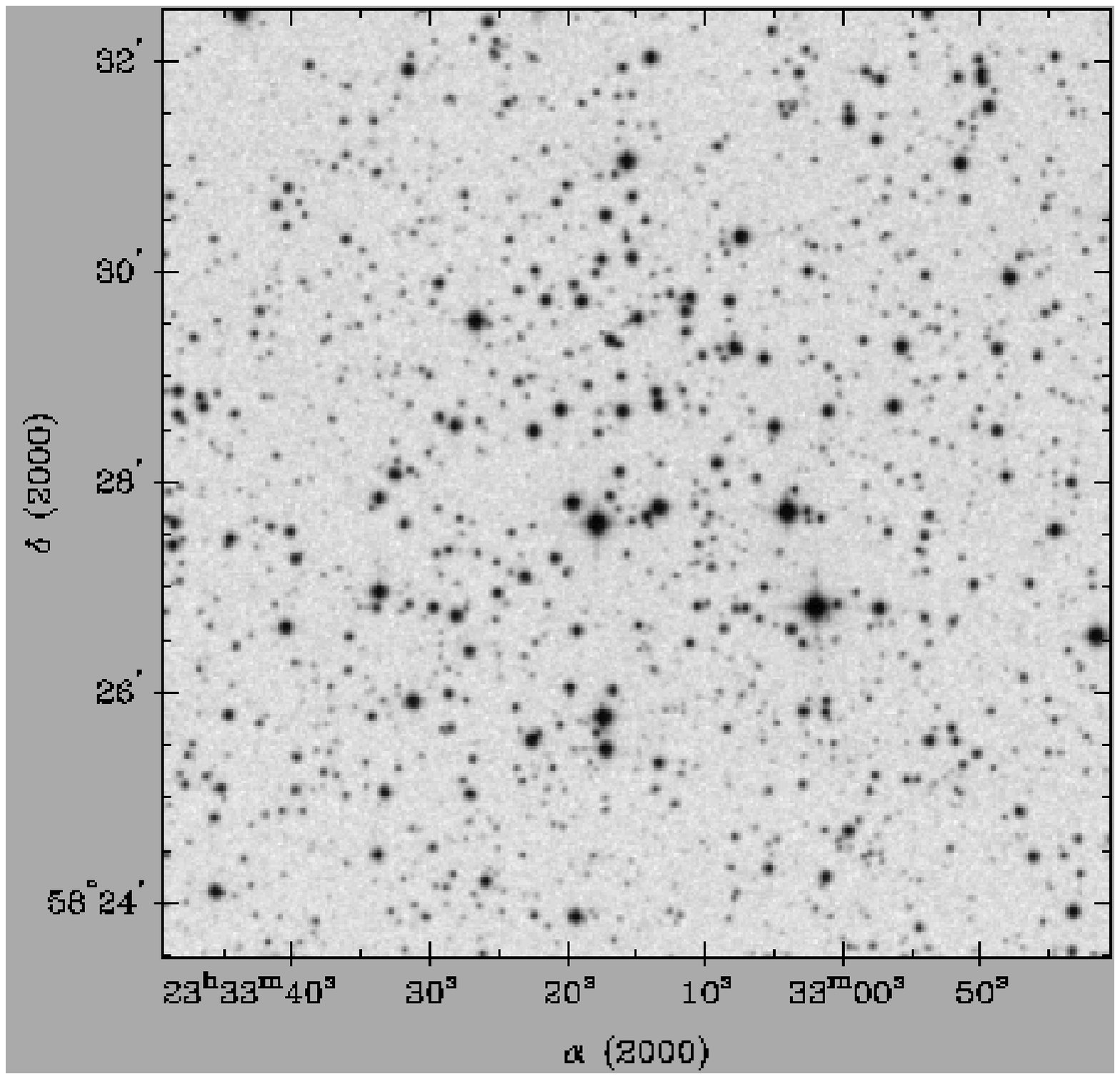}
\end{minipage}\hfill
\caption[]{Same as Fig.~\ref{fig1} for the $11\arcmin\times11\arcmin$ field around 
Lyng\aa\,12 (left panel) and the $9\arcmin\times9\arcmin$ field around King\,20 (right panel).}
\label{fig2}
\end{figure*}

Efforts to analyse unstudied OCs and derive their fundamental parameters will contribute to future 
disc studies by unveiling the properties of individual OCs themselves and constraining theories of 
molecular cloud fragmentation, star formation and stellar and dynamical evolution. These processes
will be better understood as the number of studied clusters, faint ones in particular, increases.
A side-product of more constrained parameters will be better statistics of OCs as a class.

As a first step 
in this direction we studied a series of faint OCs (Bica, Bonatto \& Dutra \cite{BBD2003}; Bonatto, 
Bica \& Dutra \cite{BBD2004}; Bica \& Bonatto \cite{BiBo2005}) using near-IR photometry. For spatial 
and photometric uniformity we work with \jj, \hh\ and \ks\ 2MASS\footnote{The Two Micron All Sky 
Survey, All Sky data release (Skrutskie et al. \cite{2mass1997}), available at {\em
http://www.ipac.caltech.edu/2mass/releases/allsky/}} Point Source Catalogue (PSC) photometry. We have 
developed methods to disentangle cluster and field stars in colour-magnitude diagrams (CMDs), as well 
as in radial stellar density distributions. The use of field-star decontamination and colour-magnitude 
filters (hereafter C-MF) have produced more robust parameters (e.g. Bonatto \& Bica \cite{BB2005}; 
Bonatto et al. \cite{BBOB06}). In particular, field-star decontamination constrains more the age, 
especially for low-latitude OCs (Bonatto et al. \cite{BKBS06}). These procedures have proven useful 
also in the analysis of faint and/or distant OCs (Bica, Bonatto \& Dutra \cite{BBD2003}; Bonatto, Bica 
\& Dutra \cite{BBD2004}; Bica \& Bonatto \cite{BiBo2005}). Recently we discussed the advantages of 
using C-MFs in the construction of radial density profiles for M\,52 and NGC\,3960, whose fields are
affected by different amounts of differential reddening (Bonatto \& Bica \cite{BB2006}). As by-products
we have been deriving more constrained structural parameters such as core and limiting radii, and mass 
function slopes, that allow inferences on dynamical evolution.

In the present work we focus on the faint OCs Lyng\aa\,2, Lyng\aa\,12, King\,20 (Alter et al. 
\cite{Alter70} and references therein), and BH\,63 that was catalogued by van den Bergh \& Hagen 
(\cite{BH75}). 

We intend to derive photometric, cluster structure and dynamic-related parameters working with a set 
of methods which provide more accurate results. The target clusters are Lyng\aa\,2 (for comparison 
purposes with the literature), and the as yet unstudied OCs Lyng\aa\,12, King\,20 and BH\,63. Since 
BH\,63 is very reddened, and consequently differential reddening may affect its field, it required 
an approach slightly different than those for the other OCs. Throughout this paper errors correspond 
to $1\sigma$ Poisson fluctuations.

\begin{table*}
\caption[]{General data on the clusters}
\label{tab1}
\tiny
\renewcommand{\tabcolsep}{0.6mm}
\renewcommand{\arraystretch}{1.5}
\begin{tabular}{lcccccccccccccc}
\hline\hline
&\multicolumn{3}{c}{Derived from XDSS}&\multicolumn{9}{c}{Derived from 2MASS}\\
\cline{2-4}\cline{6-14}
Cluster&$\alpha(2000)$&$\delta(2000)$&D&&$\alpha(2000)$&$\delta(2000)$&$\ell$&$b$&Age&$\ebv$&\ds&\dgc&\dgc&Alternative \\
&(hms)&($^\circ\arcmin\arcsec$)&(\arcmin)&&(hms)&($^\circ\arcmin\arcsec$)&($^\circ$)&($^\circ$)&(Myr)&&(kpc)&(kpc)&(kpc)&Names\\
(1)&(2)&(3)&(4)&&(5)&(6)&(7)&(8)&(9)&(10)&(11)&(12)&(13)&(14)\\
\hline
BH\,63&09:20:39&$-$49:13:23&2.5&&09:20:39.5&$-$49:13:20.5&271.63&$+$0.40&$700\pm100$&$1.9\pm0.6$&$2.3\pm0.3$
      &$8.3\pm0.3$&$7.5\pm0.3$ &ESO\,212SC2\\

Lyng\aa\,2&14:24:35&$-$61:19:50&13.0&&14:24:21.1&$-$61:19:21.0&313.84&$-0.43$&$90\pm10$&$0.22\pm0.04$
          &$0.9\pm0.1$&$7.4\pm0.1$&$6.6\pm0.1$  &OCl$-$916,\,ESO\,134SC2,\,BH\,157\\

Lyng\aa\,12&16:46:04&$-$50:45:50&8.0&&16:46:04.4&$-$50:45:40.1&335.70&$-3.46$&$560\pm100$&$0.22\pm0.04$
           &$1.0\pm0.2$&$7.1\pm0.2$&$6.3\pm0.2$    &OCl$-$973,\,ESO\,226SC22\\
 
King\,20&23:33:15&$+$58:28:00&6.0&&23:33:17.0&$+$58:28:33.1&112.85&$-$2.86&$200\pm20$&$0.65\pm0.03$
        &$1.9\pm0.2$&$8.9\pm0.2$&$8.1\pm0.2$   &OCl$-$262 \\
\hline
\end{tabular}
\begin{list}{Table Notes.}
\item Cols.~2 and 3: measured by us on XDSS images. Col.~4: angular diameter estimated on XDSS images. 
Col.~10: reddening in the cluster's central region (Sect.~\ref{age}). Col.~12: \dgc\ calculated 
using $R_O=8.0$\,kpc (Reid \cite{Reid93}) as the distance from the Sun to the Galactic center. 
Col.~13: \dgc\ using $R_O=7.2$\,kpc (Bica et al. \cite{BBBO06}).
\end{list}
\end{table*}

This paper is organized as follows. In Sect.~\ref{Target_OCs} we discuss basic properties and review 
literature data (when available) on the 4 clusters. In Sect.~\ref{2mass} we present the 2MASS data, 
subtract the field-star contamination (except for BH\,63), derive fundamental cluster parameters, and 
analyse the radial density profiles. In Sect.~\ref{MF} we derive luminosity and mass functions (LFs and 
MFs), and compute stellar content properties. In Sect.~\ref{CWODS} we discuss dynamical states. Concluding 
remarks are given in Sect.~\ref{Conclu}. 

\section{The target open clusters}
\label{Target_OCs}

Ruprecht (\cite{Ruprecht66}) classified Lyng\aa\,2 and Lyng\aa\,12 as poor loose OCs of Trumpler 
types IV2p and IV1p, respectively. King\,20 was considered somewhat more concentrated with type II1p. The 
only previously studied object is Lyng\aa\,2. Based on the photographic UBV analysis by Lindoff 
(\cite{Lindoff68}), WEBDA provides a reddening $\ebv=0.20$, distance from the Sun $\ds=1$\,kpc and 
age $\approx130$\,Myr. Kharchenko et al. (\cite{Kharchenko05}) found $\ebv=0.20$, $\ds=1$\,kpc and 
age $\approx260$\,Myr. They also estimated a core radius of $4.8\arcmin$ and a cluster radius of
$13.2\arcmin$.

In Fig.~\ref{fig1} we show XDSS\footnote{Second Generation Digitized Sky Survey, extracted from the 
Canadian Astronomy Data Centre (CADC), at \em http://cadcwww.dao.nrc.ca/} R images of BH\,63 (left panel) 
and Lyng\aa\,2 (right panel). In Fig.~\ref{fig2} we present the images of Lyng\aa\,12 (left panel) and 
King\,20 (right panel).

The clusters are projected close to the plane (Table~\ref{tab1}). Nevertheless their fields are 
differently populated owing to their Galactic longitudes. Lyng\aa\,12 and Lyng\aa\,2 are located 
in the 4th quadrant, and Lyng\aa\,12 is probably contaminated by bulge stars (Sect.~\ref{FSD}). 
They present low stellar density contrast as expected from their Trumpler classifications. 
King\,20 is in the 2nd quadrant and appears looser than a Trumpler type II cluster. Probably the 
best classification would be a type III. BH\,63, located near the border between the 3rd and 4th 
quadrants, has a significant contrast with respect to the field. It is projected in the direction of 
the dark nebula FeSt\,1-108 (Feitzinger \& St\"uwe \cite{FS84}; Dutra \& Bica \cite{DB2002}) with 
an angular diameter of 96\arcmin. The high reddening of BH\,63 (Sect.~\ref{age}) suggests that the 
nebula is located in the cluster foreground. No velocity is available for the nebula, so a distance 
estimate is not available, but the advanced dynamical state of BH\,63 (Sect.~\ref{CWODS}) might be 
related to an interaction with a molecular cloud. More detailed studies are necessary to verify this 
possibility.

In Table~\ref{tab1} we provide data on the clusters. Right ascension, declination and diameter 
(Cols.~2 to 4) were measured visually by us on XDSS images (Figs.~\ref{fig1} and \ref{fig2}) 
as the best approximation of the cluster center. However, the stellar radial density profiles  
(Sect.~\ref{struc}) presented a dip at these centers in all clusters. Consequently, we searched 
for new coordinates that would maximize the central density of stars by examining histograms for 
the number of stars in 0.5\arcmin\ bins of right ascension and declination on C-MF photometry
(Sect.~\ref{struc}). The new central coordinates and the corresponding Galactic longitude and 
latitude are given in Cols.~5 to 8 of Table~\ref{tab1}. Age, central reddening, distance from 
the Sun and Galactocentric distance based on 2MASS (Sect.~\ref{age}) are in Cols.~9 to 13. 
Alternative cluster names are in Col.~14.

\section{The 2MASS photometry}
\label{2mass}

VizieR\footnote{\em http://vizier.u-strasbg.fr/viz-bin/VizieR?-source=II/246} was used to extract 
\jj, \hh\ and \ks\ 2MASS photometry in a circular area centered on the optimized coordinates of the
4 clusters (Cols.~5 and 6 of Table~\ref{tab1}). Extraction radii are $R=20\arcmin$ (BH\,63), 
$R=35\arcmin$ (Lyng\aa\,2),
and $R=40\arcmin$ (Lyng\aa\,12 and King\,20). Our previous experience (Sect.~\ref{intro}) shows 
that as long as no other cluster is present in the field, such large extraction areas provide
the required statistics for field-star characterisation. For photometric quality 
constraints the extraction was restricted to stars {\em (i)} brighter than the 99.9\% Point 
Source Catalogue Completeness Limit\footnote{Following the Level\,1 Requirement, according to 
{\em\tiny http://www.ipac.caltech.edu/2mass/releases/allsky/doc/sec6\_5a1.html }}, $\jj=15.8$, 
$\hh=15.1$\ and $\ks=14.3$, respectively, and {\em (ii)} with errors in \jj, \hh\ and \ks\
smaller than 0.2\,mag. For reddening transformations we use the relations $A_J/A_V=0.276$, 
$A_H/A_V=0.176$, $A_{K_S}/A_V=0.118$, and $A_J=2.76\times\ejh$ (Dutra, Santiago \& Bica 
\cite{DSB2002}), assuming a constant total-to-selective absorption ratio $R_V=3.1$. 

$\jj\times\jh$ CMDs of regions that provide a suitable contrast with the field are shown in 
Fig.~\ref{fig3} (left panels). For BH\,63 we extracted photometry from a region with $R=2.5$\arcmin\ 
centered on the optimized coordinates (Table~\ref{tab1}). For Lyng\aa\,2, Lyng\aa\,12 and King\,20 
we used $R=5$\arcmin. The extension and morphology of the main sequences (MS) indicate that cluster 
age increases from Lyng\aa\,2, King\,20, Lyng\aa\,12 and BH\,63. Because of the low-latitudes, 
field stars (mostly disc) contaminate the CMDs particularly at faint magnitudes and red colours. 
Bulge stars (mostly red giants) contaminate the field of Lyng\aa\,12.

\subsection{Field-star decontamination}
\label{FSD}

\begin{figure} 
\resizebox{\hsize}{!}{\includegraphics{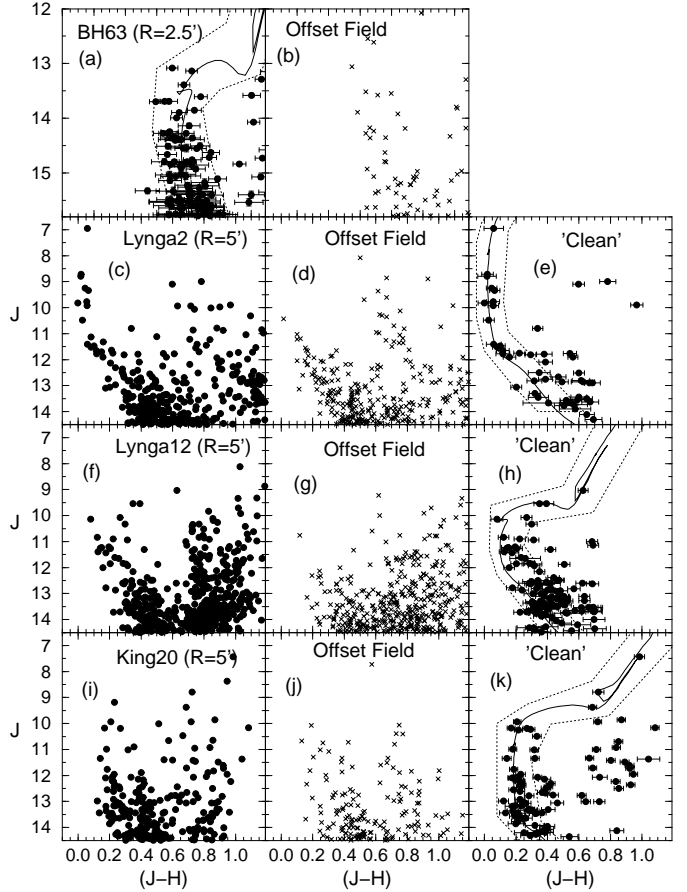}}
\caption[]{Left panels: observed CMDs. Middle panels: field stars. Right panels: field-star decontaminated 
CMDs. The solid line in panels (a), (e), (h), and (k) shows the Padova isochrone, while the dotted line
represents the C-MF used to isolate cluster MS/evolved stars. Field-star decontamination 
was not applied to BH\,63 because of high reddening.}
\label{fig3}
\end{figure}

Field stars are usually an important contaminant of CMDs, particularly of low-latitude
OCs and/or those projected against the bulge. Their presence in the fields of the target 
clusters can be seen in the corresponding offset-field CMDs (middle panels of 
Fig.~\ref{fig3}), extracted from a region outside each cluster's limiting radius (Sect.~\ref{struc})
with the same projected area as the central extraction (left panels of Fig.~\ref{fig3}). 

\begin{figure} 
\resizebox{\hsize}{!}{\includegraphics{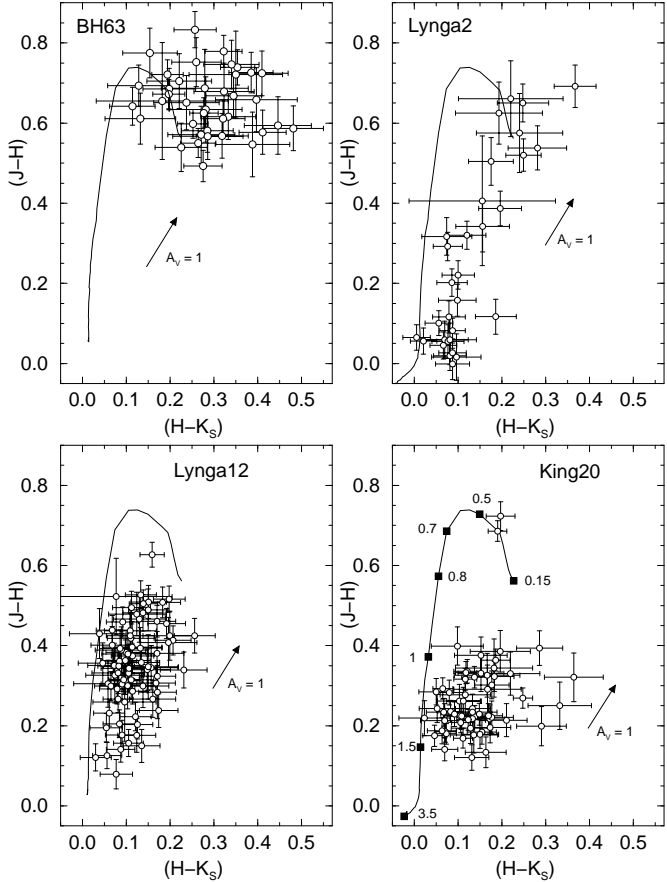}}
\caption[]{Colour-colour diagrams. Solid line: MS range of the Padova isochrones of ages 700\,Myr
(BH\,63), 90\,Myr (Lyng\aa\,2), 560\,Myr (Lyng\aa\,12), and 200\,Myr (King\,20). Arrows: 
reddening vectors $\ejh=1.72\times\ehk$ for $A_V=1$.  For illustrative purposes, representative 
masses (in \ms) are indicated along the 200\,Myr isochrone (King\,20).}
\label{fig4}
\end{figure}

To infer the intrinsic cluster-CMD morphology we use the field-star decontamination procedure
previously applied in the analysis of low-contrast (Bica \& Bonatto \cite{BiBo2005}), young 
embedded (Bonatto, Santos Jr. \& Bica \cite{BSJB05}), and young (Bonatto et al. \cite{BBOB06}) 
OCs. The algorithm works on a statistical basis that takes into account the relative densities
of stars in a cluster region and offset field. In short, it {\em (i)} computes the surface density 
of field stars (according to the number of stars in the respective offset field), {\em (ii)} assumes 
that it is uniform throughout the cluster field, {\em (iii)} divides the CMD in colour/magnitude cells 
of varying size,  {\em (iv)} randomly assigns which star within a CMD cell will be assumed to be a 
field star, and finally {\em (v)} subtracts the expected number of candidate field stars from each 
cell. The size of the colour/magnitude cells can be subsequently adjusted so that the total number 
of subtracted stars (summing up all cells) matches the expected number of field stars. 
Because the remaining stars are 
in CMD cells where the stellar density presents a clear excess over the field, they have a high 
probability of being cluster members. In crowded field regions, field-star density at faint magnitudes
(usually $\jj\geq14.5$ in 2MASS) may be equal or even larger than that expected from the cluster. In
such cases the present decontamination procedure would produce an artificially truncated MS at the sub-solar 
mass range, even for old OCs. Further details on this procedure are in Bonatto et al. (\cite{BBOB06}). 
As offset fields we take the regions $20\leq R(\arcmin)\leq35$ (Lyng\aa\,2), $25\leq R(\arcmin)\leq40$ 
(Lyng\aa\,12), and $20\leq R(\arcmin)\leq40$ (King\,20), that are large enough to produce field-star 
statistical representativity, both in magnitude and colours. This procedure could not be applied to 
BH\,63 because of the high reddening and small angular size (Sects.~\ref{age} and \ref{struc}).
Because it actually subtracts stars from the original files - thus artificially changing the 
radial distribution of stars - we use the field-star decontamination only to uncover the intrinsic
cluster CMD morphologies and build colour-colour diagrams (Fig.~\ref{fig4}). Cluster structure and
luminosity/mass functions are analysed with the stars selected with the colour-magnitude filters
shown in panels (a), (e), (h), and (k) of Fig.~\ref{fig3}.

\begin{table}
\caption[]{Cluster and field-star statistics}
\label{tab2}
\renewcommand{\tabcolsep}{0.85mm}
\renewcommand{\arraystretch}{1.2}
\begin{tabular}{lrrrrrrrrrrrr}
\hline\hline
&\multicolumn{3}{c}{Lyng\aa\,2}&&\multicolumn{3}{c}{Lyng\aa\,12}&&\multicolumn{3}{c}{King\,20}\\
\cline{2-4}\cline{6-8}\cline{10-12}
$\Delta R$&$N_{obs}$&$f_{cl}$&$f_{fs}$& &$N_{obs}$&$f_{cl}$&$f_{fs}$ & &$N_{obs}$&$f_{cl}$&$f_{fs}$\\
(\arcmin)&(stars)&(\%)&(\%)&&(stars)&(\%)&(\%)&&(stars)&(\%)&(\%)\\
   (1)   &(2)    &(3) &(4) &&(5)    &(6) &(7)&&(8) & (9)&(10)\\
\hline
0--2  &80&34&66&&167&32&68&&77&44&56\\
2--4  &1841&14&86&&442&23&77&&165&21&79\\
4--6  &271&3&97&&651&13&87&&243&11&89\\
6--8  &336&0&100&&1152&7&93&&315&4&96\\
8--10 &---&---&---&&1491&7&93&&399&0&100\\
10--12&---&---&---&&1765&4&96&&---&---&--- \\
12--14&---&---&---&&2064&3&97&&---&---&---\\
14--16&---&---&---&&2312&0&100&&---&---&---\\
\hline
Total &451&12&88&&6853&6&94&&645&18&82 \\
\hline
\end{tabular}
\begin{list}{Table Notes.}
\item Cols.~2, 5 and 8: number of observed stars in the region. Cols.~3, 6 and 9: fraction of
member stars. Cols.~4, 7 and 10: fraction of field stars. Total fields are $R=5.5\arcmin$ (Lyng\aa\,2),
$R=13\arcmin$ (Lyng\aa\,12) and $R=7\arcmin$ (King\,20).
\end{list}
\end{table}

In Table~\ref{tab2} we summarise the results of the decontamination procedure in terms of the
fraction of member and field stars throughout the fields of the target clusters. Member stars appear
to be found up to $R\approx4 - 6\arcmin$, $R\approx12 - 14\arcmin$, and $R\approx6 - 8\arcmin$ from 
the centers of Lyng\aa\,2, Lyng\aa\,12 and King\,20. Because of its near-bulge and central disc direction 
($\ell\approx336^\circ$), the fraction of member stars in the field of Lyng\aa\,12 is $\approx1/3$
of that in the field of King\,20 ($\ell\approx113^\circ$); for Lyng\aa\,2 ($\ell\approx314^\circ$)
it is intermediate between both. 

The field-star decontaminated CMDs of the $R\leq5\arcmin$ regions are shown in the right panels 
of Fig.~\ref{fig3}, where error bars are included to show the magnitude of the photometric
uncertainties. As expected, most of the red and faint stars were considered as field stars by 
the decontamination procedure. 

\subsection{Colour-colour diagrams}
\label{2CD}

Colour-colour diagrams with near-IR photometry have been useful particularly in the analysis
of young clusters, e.g. NGC\,6611 (Bonatto, Santos Jr. \& Bica \cite{BSJB05}), Trapezium and 
the cluster embedded in the nebula NGC\,2327 (Soares \& Bica \cite{SB2002}).  
We show in Fig.~\ref{fig4} the $\jh\times\hk$ diagrams of the target OCs together 
with the MS portion of the respective Padova isochrones (Sect.~\ref{age}) and the reddening 
vector ($\ejh=1.72\times\ehk$) for $A_V=1$. For membership considerations we used only the stars selected 
with the C-MFs applied to the field-star decontaminated photometry of Lyng\aa\,2, 
Lyng\aa\,12 and King\,20 (panels e, h and k of Fig.~\ref{fig3}). Because of the high reddening 
in the field of BH\,63 (Sect.~\ref{FSD}), the corresponding colour-colour diagram was produced with 
the stars selected after applying the colour-magnitude filter (panel (a) of Fig.~\ref{fig3}).
Most of the stars in Lyng\aa\,2, Lyng\aa\,12 and King\,20 distribute 
along the MS of the Padova isochrones which suggests that reddening values in their fields are not 
exceedingly high. The locus and colour-spread in BH\,63, on the other hand, reflect a high 
reddening and some differential reddening in the field.

Red stars around the M dwarf loci ($\jh\approx0.7$ and $\hk\approx0.15$) are probably residual 
field stars, since we cannot detect M dwarfs in these clusters with 2MASS. Particularly in 
Lyng\aa\,2 and Lyng\aa\,12 the red stars follow the reddening vector, again suggesting residual 
field contamination. 

\subsection{Cluster age and distance from the Sun}
\label{age}
 
Cluster age is derived with solar-metallicity - which is typical of OCs (Bonatto, Bica \& 
Girardi \cite{BBG2004}; WEBDA) - Padova isochrones (Girardi et al. 
\cite{Girardi2002}) computed with the 2MASS \jj, \hh\ and \ks\ filters\footnote{\em\tiny 
http://pleiadi.pd.astro.it/isoc\_photsys.01/isoc\_photsys.01.html. 2MASS transmission 
filters produced isochrones very similar to the Johnson-Kron-Cousins (e.g. Bessel
\& Brett \cite{BesBret88})} ones, with differences of at most 0.01 in \jh\ (Bonatto, Bica 
\& Girardi \cite{BBG2004}).

The field-star decontaminated CMD morphologies (right panels of Fig.~\ref{fig3}) provide enough 
constraints to derive the cluster ages of Lyng\aa\,2, Lyng\aa\,12 and King\,20. Particularly
for Lyng\aa\,2 and Lyng\aa\,12, the field-star decontaminated CMDs provide unambiguous isochrone
fits because of the conspicuous upper-MS in Lyng\aa\,2 and the turnoff and giant branch in Lyng\aa\,12
(panels (e) and (h) of Fig.~\ref{fig3}). To a lesser degree, the same applies to King\,20, in which 
the presence of 3 bright stars with $\jh\ge0.65$ and $\jj\le9.5$ suggest evolved stars (panel (k)).
For Lyng\aa\,2 
we found an age of $90\pm10$\,Myr, observed distance modulus $\mMJ=9.9\pm0.1$ and colour excess 
$\ejh=0.07\pm0.01$, converting to $\ebv=0.22\pm0.03$. This age-solution is plotted in panel (e) 
of Fig.~\ref{fig3}. The absolute distance modulus is $\mMo=9.7\pm0.1$, resulting in 
$\ds=0.9\pm0.1$\,kpc. These values are in good agreement with those in the literature 
(Sect.~\ref{Target_OCs}). The turnoff occurs at $M_J\approx-2.1$ and $m\approx5.4\ms$. Because 
of the 2MASS faint-magnitude limit (Sect.~\ref{2mass}), MS stars are detected for $m\geq0.9\,\ms$.  
The Galactocentric distance of Lyng\aa\,2 is $\dgc=7.4\pm0.2$\,kpc, using $R_O=8.0$\,kpc as the 
Sun's distance to the Galactic center (Reid \cite{Reid93}). However, with the recently derived value 
of $R_O=7.2$\,kpc (based on updated parameters of globular clusters - Bica et al. \cite{BBBO06}), 
this OC ends up located $\dgc=6.6\pm0.1$\,kpc from the Galactic center. In any case, Lyng\aa\,2 is 
located $\approx0.6$\,kpc inside the Solar circle. 

For Lyng\aa\,12 we derived $\ejh=0.07\pm0.01$ ($\ebv=0.22\pm0.03$), an age of $560\pm100$\,Myr and
$\ds=1.0\pm0.1$\,kpc (panel h); MS stars are detected in the range $0.8\leq m(\ms)\leq2.5$. For 
King\,20 we found $\ejh=0.21\pm0.01$ ($\ebv=0.65\pm0.03$), an age of $200\pm20$\,Myr and 
$\ds=1.9\pm0.2$\,kpc (panel k). MS stars are detected in the range $1.2\leq m(\ms)\leq3.8$. Additional 
parameters are given in Table~\ref{tab1}.

Parameters of BH\,63 were based on the observed photometry. We found $\ejh=0.60\pm0.02$ 
($\ebv=1.9\pm0.6$), ${\rm age}=700\pm100$\,Myr and $\ds=2.3\pm0.3$\,kpc; the detected MS mass 
range is $1.5\leq m(\ms)\leq2.3$ (panel a). BH\,63 is a heavily reddened cluster, with
$A_V\approx6.1$.

\subsection{Cluster structure}
\label{struc}

Cluster structure was inferred with the radial density profile (RDP), defined as the projected
number-density of MS/evolved stars around the center. The RDPs were built with stars selected 
with the C-MFs (right panels of Fig.~\ref{fig3}). The use 
of C-MFs to discard foreground/background field stars was applied 
in the analysis of the OCs M\,67 (Bonatto \& Bica \cite{BB2003}), NGC\,3680 (Bonatto, Bica \& Pavani
\cite{BBP2004}), NGC\,188 (Bonatto, Bica \& Santos Jr. \cite{BBS2005}), NGC\,6611 (Bonatto, Santos 
Jr. \& Bica \cite{BSJB05}), and NGC\,4755 (Bonatto et al. \cite{BBOB06}). To avoid oversampling near 
the center and undersampling for large radii, the RDPs were built by counting stars in concentric rings 
with radius $\Delta R=0.5\arcmin$ for $0\leq R(\arcmin)<5$, $\Delta R=1\arcmin$ for $5\leq 
R(\arcmin)<10$, $\Delta R=2\arcmin$ for $10\leq R(\arcmin)<30$ and $\Delta R=4\arcmin$ for 
$R\geq30\arcmin$. 

\begin{figure} 
\resizebox{\hsize}{!}{\includegraphics{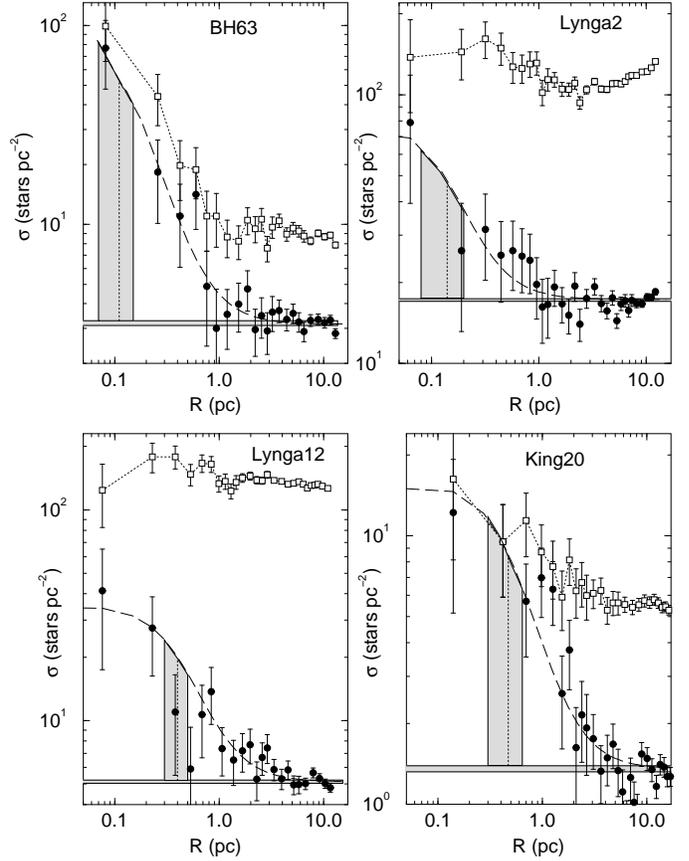}}
\caption[]{Stellar radial density profiles. Filled circles: colour-magnitude filtered RDP. Dashed line: 
best-fit two-parameter King profile. Horizontal shaded region: residual stellar background level 
measured in the colour-magnitude filtered offset field. Vertical dotted line: best-fit King core 
radius. Vertical shaded region: standard deviation of the core radius. For comparison purposes the empty 
squares show the RDPs prior to the C-MF, where the actual stellar background level can be
seen.}
\label{fig5}
\end{figure}

Fig.~\ref{fig5} shows the RDPs for the MS/evolved stars of the present OCs. For absolute comparison between clusters 
the radius scale was converted to parsecs and the number-density of stars to $\rm stars\,pc^{-2}$\ using the 
distances derived in Sect.~\ref{age}. We also show in Fig.~\ref{fig5} the observed RDPs produced with photometry 
prior to the use of C-MF. Clearly, the filtered profiles present less fluctuations and go deeper into the cluster 
structure than the observed ones. In particular, the observed profiles tend to underestimate the cluster's extension.
As expected, the C-MF produced residual background levels much lower than the observed ones, especially
for Lyng\aa\,2 and Lyng\aa\,12.
Besides, C-MFs have proven to be essential to unveil the cluster centroid especially for faint OCs such as those 
in the present sample.

\begin{table*}
\caption[]{Structural parameters.}
\label{tab3}
\renewcommand{\tabcolsep}{1.85mm}
\renewcommand{\arraystretch}{1.3}
\begin{tabular}{lccccccccccc}
\hline\hline
Cluster&1\arcmin&$\sigma_{bg}$&$\sigma_{0K}$&\rc&\rl&&$\sigma_{bg}$&$\sigma_{0K}$&\rc&\rl \\
       &(pc)&$\rm(stars\,\arcmin^{-2})$&$\rm(stars\,\arcmin^{-2})$&(\arcmin)&(\arcmin)&
       &$\rm(stars\,pc^{-2})$&$\rm(stars\,pc^{-2})$&(pc)&(pc)\\
(1)&(2)&(3)&(4)&(5)&(6)&&(7)&(8)&(9)&(10)\\
\hline
BH\,63&0.680&$1.46\pm0.03$&$52\pm29$&$0.16\pm0.06$&$2.4\pm0.5$ &
      &$3.1\pm0.1$&$112\pm63$&$0.11\pm0.04$&$1.6\pm0.3$ \\
Lyng\aa\,2&0.254&$1.12\pm0.01$&$4.1\pm2.2$&$0.54\pm0.24$&$5.5\pm0.8$ &
          &$17.3\pm0.2$&$63\pm34$&$0.14\pm0.06$&$1.4\pm0.2$\\
Lyng\aa\,12&0.304&$0.48\pm0.01$&$2.8\pm1.2$&$1.30\pm0.37$&$13\pm2$ &
           &$4.8\pm0.1$&$30\pm13$&$0.40\pm0.10$&$3.9\pm0.6$ \\
King\,20&0.560&$0.43\pm0.01$&$4.5\pm2.6$&$0.84\pm0.33$&$7\pm1$ &
        &$1.4\pm0.1$&$14\pm8$&$0.47\pm0.18$&$3.9\pm0.6$ \\
\hline
\end{tabular}
\begin{list}{Table Notes.}
\item Col.~2: arcmin to parsec scale. We express King profile as 
$\sigma(R)=\sigma_{bg}+\sigma_{0K}/(1+(R/R_{\rm core})^2)$. To minimise degrees of freedom 
in the fit $\sigma_{bg}$ was kept fixed (measured in the respective offset fields) while 
$\sigma_{0K}$ and \rc\ were allowed to vary. Note that $\sigma_{0K}$ and $\sigma_{bg}$
represent residual densities of stars, measured on the C-MF photometry.
\end{list}
\end{table*}

Structural parameters were derived by fitting the C-MF RDPs with the two-parameter 
King (\cite{King1966a}) surface density profile, which describes the central and intermediate regions of 
normal clusters (King \cite{King1966b}; Trager, King \& Djorgovski \cite{TKD95}). The fits were performed 
using a nonlinear least-squares fit routine that uses errors as weights. To minimise degrees of freedom
in the fit the background level ($\sigma_{bg}$) was kept constant, corresponding to the residual values
measured in the corresponding offset fields (Sect.~\ref{FSD}). Parameters derived from the fit are the
King
central density of stars ($\sigma_{0K}$) and core radius (\rc). The resulting parameters are given in 
Table~\ref{tab3} and the best-fit solutions are superimposed on the C-MF RDPs (Fig.~\ref{fig5}). A limiting 
radius (\rl) of a cluster can be estimated by considering the fluctuations of the RDPs with respect to the 
 residual background. \rl\ describes where the RDP merges into the background and, for practical 
purposes most of the 
cluster stars are contained within $\rl$. For comparison purposes Table~\ref{tab3} provides parameters in 
angular and absolute units. Probably because of different methods and data sets the present values of \rc\ 
and \rl\ for Lyng\aa\,2 correspond to about $1/9$ and $1/2$ of those given in Kharchenko et al. 
(\cite{Kharchenko05}). We verified that the centroid in both works is virtually coincident. The difference 
may be attributed to their brighter limits (Kharchenko et al. \cite{KPRSS04}) producing shallower profiles 
for this faint cluster.

Compared to other OCs (Sect.~\ref{CWODS}; Nilakshi \cite{Nilakshi2002}; Tadross et al. \cite{Tad2002}), 
the present clusters are small both in core and limiting radii. This is particularly true for Lyng\aa\,2 
and BH\,63.

Within uncertainties, King profiles provide a good analytical representation of the stellar RDPs of 
Lyng\aa\,12 and King\,20, from the external parts to the core. However, because of the small size of 
BH\,63 and Lyng\aa\,2 (and the larger distance of BH\,63) their RDPs do not reach much into the core. 
In BH\,63 (and to a lesser extent Lyng\aa\,2) the RDP seems to increase almost linearly (in a log-log 
scale) with decreasing distance to the center with no evidence of the King-like core turnover. Consequently, 
we cannot compare the actual stellar density profiles of BH\,63 and Lyng\aa\,2 inside the core with the 
King profile. Two interpretations are possible. Either the very central region was not accessed or the 
linear increase points to a post-core collapse structure as those observed by Trager, King \& Djorgovski 
(\cite{TKD95}) for several post-core collapse globular clusters. 

Since it follows from an isothermal (virialized) sphere, the close similarity of a cluster's stellar 
RDP with a King profile suggests that the internal structure (particularly the core) has reached some 
important level of energy equipartition at the cluster age. We return to this point in Sect.~\ref{CWODS}. 

\section{Luminosity and mass functions}
\label{MF}

The 2MASS quality photometry along with the methods presented in Bonatto \& Bica 
(\cite{BB2005} and references therein) allows one to derive luminosity and mass functions 
$\left(\phi(m)=\frac{dN}{dm}\right)$ for the core, halo and overall (i.e. the whole 
cluster) regions for the present faint clusters. 

\begin{figure*}
\begin{minipage}[b]{0.50\linewidth}
\includegraphics[width=\textwidth]{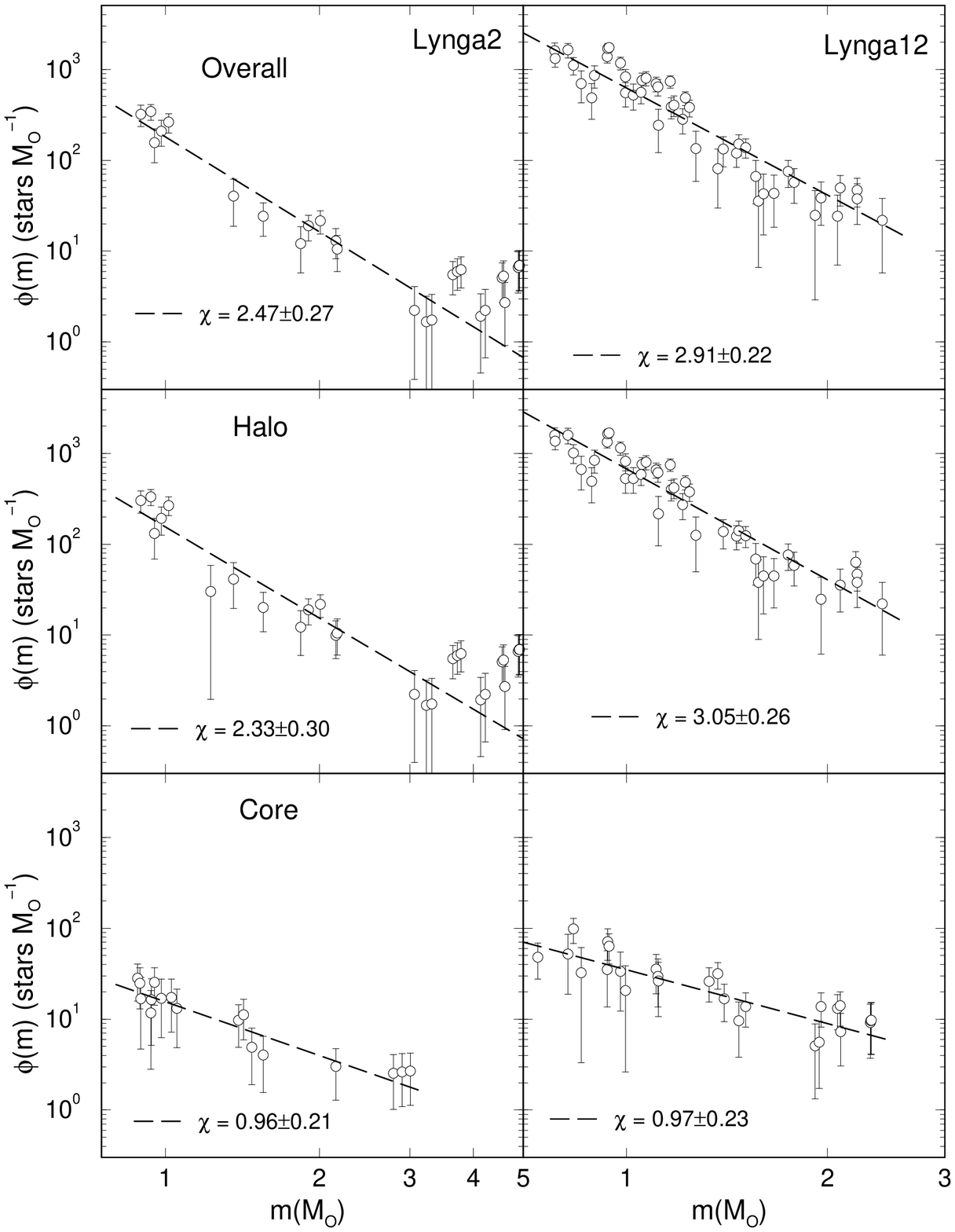}
\end{minipage}\hfill
\begin{minipage}[b]{0.50\linewidth}
\includegraphics[width=\textwidth]{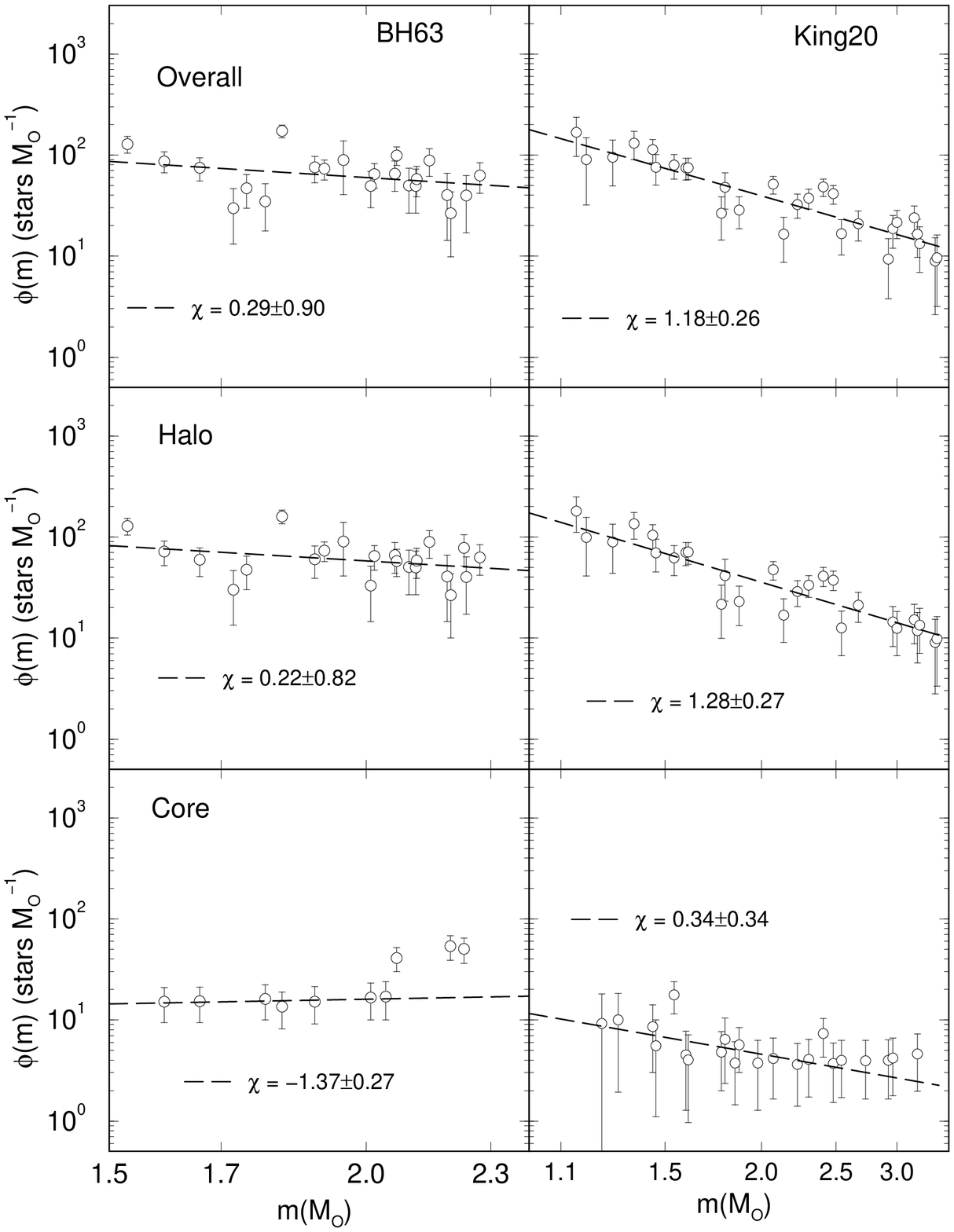}
\end{minipage}\hfill
\caption[]{Colour-magnitude filtered mass functions (empty circles) fitted with the function 
$\phi(m)\propto m^{-(1+\chi)}$ (dashed line). Top, middle and bottom panels contain the MFs of
the overall cluster, halo and core regions.}
\label{fig6}
\end{figure*}

\begin{table*}
\caption[]{Parameters related to mass functions and dynamical states}
\label{tab4}
\renewcommand{\tabcolsep}{1.15mm}
\renewcommand{\arraystretch}{1.2}
\begin{tabular}{ccccccccccccc}
\hline\hline
&\multicolumn{2}{c}{Evolved}&&\multicolumn{3}{c}{Observed MS}&&
\multicolumn{5}{c}{$\rm Evolved\ +\ Extrapolated\ MS\ $}\\
\cline{2-3}\cline{5-7}\cline{9-13}
Region&$N^*$&$m$&&$\chi$&$N^*$&\mobs&&$N^*$&$m$&$\sigma$&$\rho$&$\tau$\\
&(stars)&(\ms)&&&(stars)&(\ms)&& ($10^2$stars)&($10^2\ms$)&($\rm \ms\,pc^{-2}$)&($\rm \ms\,pc^{-3}$)\\
 (1)  & (2)   & (3)  &&(4)     & (5)    & (6)   && (7)    & (8)   &(9) &(10) & (11)\\
\hline
&\multicolumn{12}{c}{BH\,63 --- MS: $1.5\leq m(\ms)\leq2.3$ --- $\rm{Age}=700\pm100$\,Myr}\\
\cline{2-13}
Core& --- & --- &&$-1.37\pm0.27$& $12\pm3$&$22\pm5$
&&$0.3\pm0.1$&$0.4\pm0.1$&$960\pm150$&$6500\pm1000$&$19200\pm7690$\\

Halo&$2\pm1$ &$5\pm2$ &&$0.22\pm0.82$& $46\pm33$&$87\pm62$
 &&$5.6\pm5.3$&$3.1\pm1.4$&$38\pm17$&$18\pm8$&---\\

Overall& $2\pm1$&$5\pm2$ &&$0.29\pm0.90$& $48\pm25$&$90\pm47$ 
&&$6.5\pm6.0$&$3.4\pm1.2$&$42\pm15$&$20\pm7$&$108\pm88$\\
\cline{2-13}
&\multicolumn{12}{c}{Lyng\aa\,2 --- MS: $0.9\leq m(\ms)\leq5.4$ --- $\rm{Age}=90\pm10$\,Myr}\\
\cline{2-13}

Core& --- & --- &&$0.96\pm0.21$& $12\pm2$&$19\pm3$
&&$1.0\pm0.7$&$0.4\pm0.1$&$612\pm182$&$3060\pm909$&$697\pm475$\\

Halo&$1\pm1$ &$5\pm5$ &&$2.33\pm0.30$& $83\pm16$&$119\pm23$
 &&$13\pm10$&$4.6\pm1.9$&$75\pm31$&$40\pm16$&---\\

Overall& $1\pm1$&$5\pm5$ &&$2.47\pm0.27$& $94\pm17$&$136\pm27$ 
&&$15\pm11$&$5.3\pm2.2$&$86\pm36$&$46\pm19$&$7.7\pm5.3$\\ 

\cline{2-13}
&\multicolumn{12}{c}{Lyng\aa\,12 --- MS: $0.8\leq m(\ms)\leq2.5$ --- $\rm{Age}=560\pm100$\,Myr}\\
\cline{2-13}
Core& $1\pm1$ & $3\pm3$ &&$0.97\pm0.23$&$33\pm3$&$42\pm5$
&&$2.2\pm1.6$&$1.0\pm0.3$&$187\pm58$&$350\pm110$&$846\pm534$\\

Halo&$7\pm3$ &$18\pm8$ &&$3.05\pm0.26$& $539\pm37$&$556\pm40$
 &&$83\pm65$&$25\pm12$&$54\pm25$&$10\pm5$&---\\

Overall& $8\pm3$&$20\pm8$ &&$2.91\pm0.22$& $500\pm31$&$521\pm34$ 
&&$74\pm58$&$23\pm11$&$48\pm22$&$9.2\pm0.4$&$4.3\pm3.1$\\
\cline{2-13}
&\multicolumn{12}{c}{King\,20 --- MS: $1.1\leq m(\ms)\leq3.8$ --- $\rm{Age}=200\pm20$\,Myr}\\
\cline{2-13}

Core& $1\pm1$ & $3\pm3$ &&$0.34\pm0.34$& $10\pm3$&$20\pm6$
&&$0.6\pm0.2$&$0.4\pm0.1$&$56\pm12$&$90\pm20$&$735\pm365$\\

Halo&$6\pm2$ &$20\pm7$ &&$1.28\pm0.27$& $79\pm20$&$135\pm40$
 &&$13\pm9$&$5.5\pm1.9$&$12\pm4$&$2.2\pm0.7$&---\\

Overall& $7\pm3$&$24\pm10$ &&$1.18\pm0.26$& $91\pm22$&$157\pm44$ 
&&$13\pm9$&$5.6\pm1.8$&$12\pm4$&$2.2\pm0.7$&$7.0\pm4.5$\\ 

\hline\hline
\end{tabular}
\begin{list}{Table Notes.}
\item Col.~6: stellar mass of the observed MS. Col.~8: mass extrapolated to 0.08\,\ms.
Col.~11: dynamical-evolution parameter $\tau={\rm age}/t_{\rm rel}$. 
\end{list}
\end{table*}

In all clusters, the resulting faint-magnitude limit of the MS stars, isolated with the 
C-MFs (Fig.~\ref{fig3}), is brighter than that of the 99.9\% Completeness Limit 
(Sect.~\ref{2mass}). To take the residual field-star contamination into account, we build LFs 
for each cluster region and 
offset field separately. \jj, \hh\ and \ks\ LFs are built by counting stars in magnitude bins from the 
respective faint magnitude limit to the turn-off, for cluster and offset field regions. Magnitude 
bins are wider in the upper MS than in the lower MS to avoid undersampling near the turn-off and 
oversampling at the faint limit. Corrections are made for different solid angles between offset 
field and cluster regions. Cluster LFs are obtained by subtracting the offset-field LFs. They are 
transformed into MFs using the mass-luminosity relations obtained from the respective Padova isochrones 
and observed distance modulii (Sect.~\ref{2mass}).  These procedures are applied independently to the 
three 2MASS bands. Because of differences among the \jj, \hh, and \ks\ mass-luminosity relations, the
respective MFs are sampled in different mass values. Consequently, the final MF is simply a superposition 
of the \jj, \hh\ and \ks\ MFs. Detected MS mass ranges are $1.5\leq m(\ms)\leq2.3$, $0.9\leq m(\ms)\leq5.4$,
$0.8\leq m(\ms)\leq2.5$ and $1.1\leq m(\ms)\leq3.8$, 
respectively for BH\,63, Lyng\aa\,2, Lyng\aa\,12 and King\,20. Sub-solar mass stars are expected to 
populate the MS of OCs older than $\approx100$\,Myr, however, they cannot be detected with 2MASS 
because of the increasing fraction of field-star contamination in that mass range (right panels of 
Fig.~\ref{fig3}). The observed MS range was fitted with the function $\phi(m)\propto m^{-(1+\chi)}$.

The overall, halo and core MFs of Lyng\aa\,2 and Lyng\aa\,12 are shown in the left panels of
Fig.~\ref{fig6}, while those of BH\,63 and King\,20 are in the right panels. The MF slopes are 
given in col.~4 of Table~\ref{tab4}. 

A common feature among the OCs is the flat core MF, particularly in BH\,63; the overall MF of this
OC is much flatter than Salpeter's (\cite{Salpeter55}) IMF ($\chi=1.35$). Lyng\aa\,2 and Lyng\aa\,12
have halo and overall MFs steeper than Salpeter (Sect.~\ref{CWODS}).

\subsection{Cluster mass}
\label{TM}

Parameters derived from the core, halo and overall MFs of the target clusters are given in Table~\ref{tab4}. 
The number of evolved stars (col.~2) in each region was obtained by integration of the respective field-star 
subtracted LF for stars brighter than the turnoff. Multiplying this number by the mass at the turn-off yields 
an estimate of the mass stored in evolved stars (col.~3). The observed number of MS stars and corresponding 
mass (cols.~5 and 6, respectively) were derived by integrating the MFs in the mass ranges. 

To estimate the total mass locked up in stars we extrapolate the observed MFs down to the H-burning mass 
limit ($0.08\,\ms$) since these stars are expected to be present in the MS of OCs older than 90\,Myr. In
the mass range below the present detection threshold (Table~\ref{tab4}) we base the extrapolation on 
Kroupa's (\cite{Kroupa2001}) universal Initial Mass Function (IMF), in which $\chi=0.3\pm0.5$ for the range 
$0.08\leq m(\ms)\leq0.5$ and $\chi=1.3\pm0.3$ for $0.5\leq m(\ms)\leq1.0$. We remark that in the cases where 
our MF slopes are flatter than or similar (within uncertainties) to Kroupa's, we directly use the former slopes,
e.g. the three MFs of BH\,63; for the core MF of Lyng\aa\,2 we use our slope in the range $0.5\leq m(\ms)\leq0.9$,
and Kroupa's for $0.08\leq m(\ms)\leq0.5$.

The total (extrapolated MS $+$ evolved) values of number, mass, projected and volume densities 
are given in cols.~7 to 10 of Table~\ref{tab4}.

\section{Diagnostic-diagrams of dynamical states}
\label{CWODS}

The King-like RDPs (Fig.~\ref{fig5}) and the change in MF slope from core to halo (Fig.~\ref{fig6})
may be a consequence of dynamical evolution in the present OCs, particularly in the core.
At this point it is interesting to see how these faint clusters compare with a set of nearby
OCs in terms of structure and dynamical state.

\begin{figure} 
\resizebox{\hsize}{!}{\includegraphics{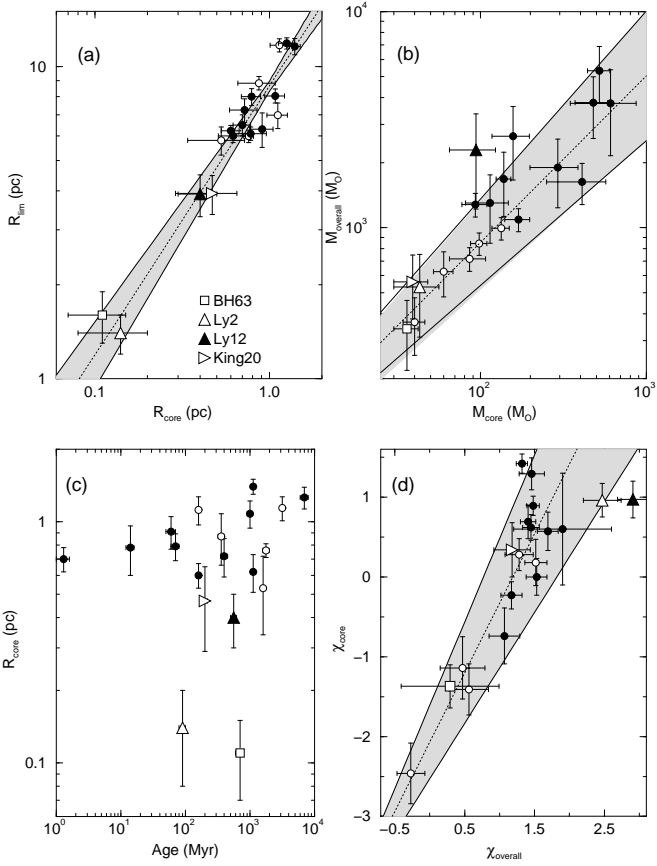}}
\caption[]{Relations involving core and limiting radii (panel a), core and overall mass (panel b),
core radius and cluster age (panel c), and core and overall MF slope (panel d). Filled symbols: 
clusters more massive than 1\,000\,\ms\ (Bonatto \& Bica \cite{BB2005}). Open symbols: 
$m<1\,000$\,\ms. Dotted lines: least-squares fits to the reference clusters. Shaded areas: 
$1\sigma$ borders of the least-squares fits. }
\label{fig7}
\end{figure}
 
\begin{figure} 
\resizebox{\hsize}{!}{\includegraphics{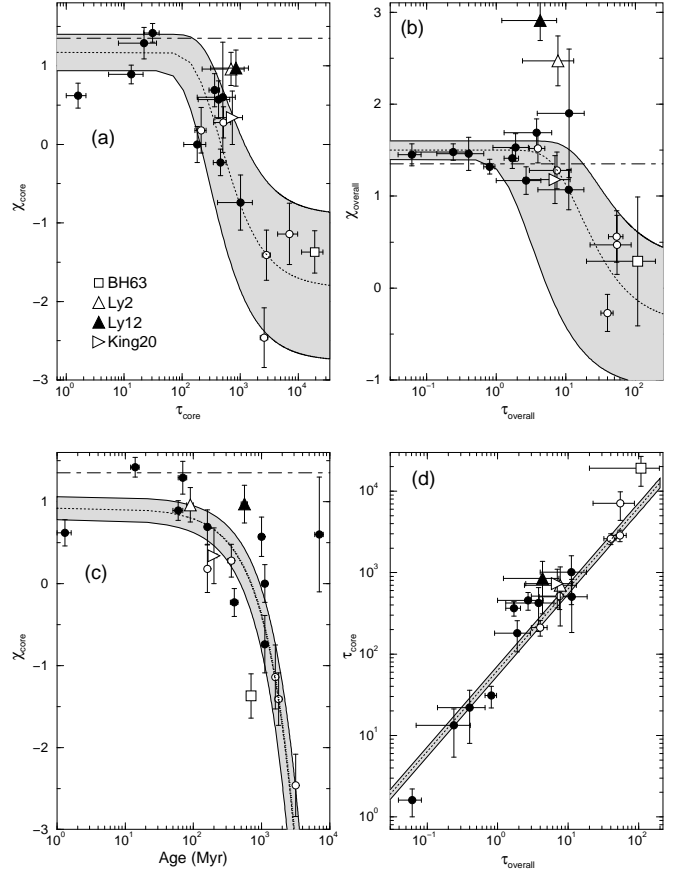}}
\caption[]{Relations involving MF slope and evolutionary parameter for the core (panel a) and
overall cluster (panel b); core MF slope and cluster age (panel c); and core and overall evolutionary 
parameter (panel d). The dot-dash line in panels (a), (b) and (c) indicates the Salpeter IMF slope 
value. Symbols as in Fig.~\ref{fig7}.}
\label{fig8}
\end{figure}

Bonatto \& Bica (\cite{BB2005}) derived a set of parameters related to the structure and
dynamical state of nearby OCs with ages in the range $70-7\,000$\,Myr and masses within 
$400-5\,300$\,\ms, following most of the present methodology. To the original sample were 
added the very young OC (age $\sim1.3$\,Myr) NGC\,6611 (Bonatto, Santos Jr. \& Bica \cite{BSJB05}) 
and the young one (age $\sim14$\,Myr) NGC\,4755 (Bonatto et al. \cite{BBOB06}). The evolutionary 
parameter $\tau=\rm{age}/\tr$ (col.~11 of Table~\ref{tab4}), where \tr\ is the relaxation time,
was found to be a good tracer of dynamical states.  In general terms, $\tau$ is a measure of the 
dynamical age of a cluster as a whole or of a cluster's internal region. In this sense, effects of
dynamical processes such as mass segregation and low-mass star evaporation on MF slopes are expected 
to become more conspicuous with increasing $\tau$. In particular, Bonatto \& Bica (\cite{BB2005})
observed that significant flattening in core and overall MFs due to dynamical effects occurs for 
$\tau_{\rm core}\geq100$ and $\tau_{\rm overall}\geq7$, respectively. The difference in core and
overall $\tau$ threshold is associated with the different average physical conditions in both 
regions, such as density and gravitational potential, which in turn affect \tr. Further discussions 
and details on parameter correlation in the reference sample are in Bonatto \& Bica (\cite{BB2005}).

Core and overall radii in the reference sample are related by the equation 
$\rl\propto R_{\rm core}^{(0.9\pm0.1)}$. Despite the small sizes of BH\,63 and Lyng\aa\,2, the present 
OCs have \rl\ and \rc\ consistent with that relation at the $1\sigma$ level (panel a of Fig.~\ref{fig7}). 
The same occurs for the core and overall (extrapolated) mass ($m_{overall}\propto m_{core}^{(0.8\pm0.1)}$, 
panel b). Within uncertainties, both slopes are the same which is probably a consequence of scaling, in 
the sense that on average massive clusters tend to be big. About $2/3$ of the reference sample suggests a 
trend of increasing core radius with age (panel c), while the remaining ones - Lyng\aa\,2 and BH\,63 in 
particular - appear to follow a sequence of decreasing \rc\ with age. Except for the steep overall MF of 
Lyng\aa\,12, the remaining OCs have MF slopes consistent with those of the reference sample (panel d).

Relations involving MF slope and dynamical-evolution parameter (Fig.~\ref{fig8}) can be used for 
inferences on dynamical states. Panels (a) and (b) suggest that both the core and overall MF slopes decay
with the evolutionary parameter according to $\chi(\tau)=\chi_o-\chi_1\times\,e^{-\left(\frac{\tau_o}
{\tau}\right)}$ (Bonatto \& Bica \cite{BB2005}), where $\chi_o$ represents the MF slope at cluster
birth, while $\chi_1$ gives the amount of slope flattening for advanced evolutionary states. Panel 
(c) suggests a systematic flattening of the core 
MF slope with cluster age, particularly for the less-massive OCs\footnote{The deviant object is the 
$\approx7$\,Gyr, massive OC NGC\,188 whose orbit avoids the inner regions of the Galaxy for most of the 
time (Bonatto, Bica \& Santos Jr. \cite{BBS2005} and references therein).}. Except for NGC\,188, the age 
dependence of the core MF slope can be parameterised by the linear-decay function $\chi(t)=\chi_o-t/t_f$, 
where $\chi_o$ may be interpreted as the core MF slope in the early phases of the cluster, and $t_f$ is the 
flattening time scale associated mostly with mass segregation. We find $t_f=775\pm130$\,Myr and 
$\chi_o=0.92\pm0.14$, a value significantly flatter than that of Salpeter IMF. Core and overall evolutionary
parameters (panel d) are related by $\tau_{overall}\propto\tau_{core}^{(0.9\pm0.1)}$. Further discussions are 
in Bonatto \& Bica (\cite{BB2005}). 

As shown in all panels of Fig.~\ref{fig8}, BH\,63 presents clear signs of  accelerated dynamical evolution, 
particularly in the core. With flat MF slopes and large dynamical evolution parameters, the core and 
overall regions of BH\,63 appear to be extreme cases of OC dynamical evolution (especially in panels a 
and b). One possibility is that interaction with the nebula FeSt\,1-108 (Sect.~\ref{Target_OCs}) may have 
accelerated its dynamical evolution. Although to a  lesser degree, similar evidence for the core and overall 
regions of King\,20 is provided in all panels. The cores of Lyng\aa\,2 and Lyng\aa\,12 show marginal 
evidence of dynamical evolution (flat MF slopes).
 
\section{Concluding remarks}
\label{Conclu}

In this paper, we derived photometric and structural parameters of the faint open clusters BH\,63, 
Lyng\aa\,2, Lyng\aa\,12, and King\,20 with \jj, \hh\ and \ks\ 2MASS photometry, restricted to stars 
with observational uncertainties $\epsilon_{J,H,K_S}<0.2$\,mag. BH\,63 is significantly more reddened
than the others, with $\ebv\approx1.9$. Distance from the Sun ranges from $\ds\approx0.9$\,kpc (Lyng\aa\,2)
to $\ds\approx2.3$\,kpc (BH\,63). Lyng\aa\,2 and Lyng\aa\,12 are located inside the Solar circle.

The use of C-MFs allowed us to obtain accurate parameters for such faint clusters
including mass function slopes for the core and halo. In particular, radial density profiles built with 
C-MF photometry are better defined and yield more constrained structural parameters 
than the observed ones. The RDPs of Lyng\aa\,12 and King\,20 are well-represented by a King profile. 
Lyng\aa\,2 and BH\,63 appear to be very small, with core and limiting radii of $\approx0.12$\,pc and 
$\approx1.5$\,pc. However, they fit in the small core and limiting radii tail of the OC distribution. 
Lyng\aa\,12 and King\,20 have $\rc\approx0.43$\,pc and $\rl\approx3.9$\,pc.

The open clusters of the present sample have flat core MFs, suggesting dynamically evolved systems, 
particularly BH\,63, which has a flat MF in  its halo as well. The total mass locked up in stars
(extrapolated down to $0.08$\,\ms) ranges from $\approx340$\,\ms\ (BH\,63) to $\approx2300$\,\ms\ 
(Lyng\aa\,12).

We also analysed the clusters with dynamical and structural diagnostic diagrams. BH\,63 appears to be 
an extreme case of OC dynamical evolution both in  its core and halo, perhaps due to interaction 
with the nebula FeSt\,1-108. King\,20 shows signs of dynamical evolution as well, but to a  lesser
degree. The cores of Lyng\aa\,2 and Lyng\aa\,12 present marginal evidence of dynamical evolution.

The present work shows that it is possible to explore faint OCs with 2MASS photometry, provided 
care is taken to identify probable member stars (to better define the cluster sequence on CMDs) 
and discard stars with discordant colour (for more intrinsic cluster RDPs and MFs). The use of
field-star decontamination and colour-magnitude filters on the fields of BH\,63, Lyng\aa\,2, 
Lyng\aa\,12 and King\,20 allowed us to obtain  good results for the fundamental, 
structural and dynamical-evolution related parameters.

\begin{acknowledgements}
We thank the anonymous referee for helpful suggestions.
This publication makes use of data products from the Two Micron All Sky Survey, which 
is a joint project of the University of Massachusetts and the Infrared Processing and 
Analysis Center/California Institute of Technology, funded by the National Aeronautics 
and Space Administration and the National Science Foundation. This research has made use 
of the WEBDA database, operated at the Institute for Astronomy of the University of Vienna.
CB and EB acknowledge support from the Brazilian Institution CNPq, and RB a CNPq/UFRGS
PIBIC fellowship.
\end{acknowledgements}

%

\end{document}